\documentclass{IEEEtran}

\usepackage{cite}
\usepackage{amsmath,amssymb,amsfonts}
\usepackage{algorithmic}
\usepackage{graphicx}
\usepackage{textcomp}

\usepackage{media9}

\usepackage{multirow}
\usepackage{epsfig}
\usepackage{graphics}

\usepackage{amsmath}
\usepackage{url}
\usepackage{array,tabularx}
\usepackage{cite}  
\usepackage{graphics} 
\usepackage{epsfig} 
\usepackage{mathptmx} 
\usepackage{times} 
\usepackage{amssymb}  
\usepackage{xfrac}
\usepackage{comment}
\usepackage{epstopdf}
\usepackage{bm}
\usepackage[export]{adjustbox}
\usepackage{multimedia}
\usepackage{xcolor}

\newcommand\x{6.3}

\definecolor{matlab_purple}{RGB}{0.72, 0.27, 1}

\usepackage{calligra}
\usepackage[T1]{fontenc}

\usepackage[english]{babel}

\usepackage[latin1]{inputenc}

\usepackage{times}

\usepackage{comment}
\usepackage{color}

\def\BibTeX{{\rm B\kern-.05em{\sc i\kern-.025em b}\kern-.08em
    T\kern-.1667em\lower.7ex\hbox{E}\kern-.125emX}}

\begin{document}

\title{
Temporal negative refraction}

\author{Or Lasri and Lea Sirota
\thanks{Or Lasri is with the School of Mechanical Engineering,
Tel Aviv University, Tel-Aviv 69978, Israel (e-mail: orlasri@mail.tau.ac.il)}
\thanks{Lea Sirota is with the School of Mechanical Engineering,
Tel Aviv University, Tel-Aviv 69978, Israel (e-mail: leabeilkin@tauex.tau.ac.il). Corresponding author.}}

\maketitle

\IEEEtitleabstractindextext{%

\begin{abstract}

Negative refraction is a peculiar wave propagation phenomenon that occurs when a wave 
crosses a boundary between a regular medium and a medium with both constitutive parameters negative at the given frequency. 
The phase and group velocities of the transmitted wave then turn
anti-parallel.
Here we propose a temporal analogue of the negative refraction phenomenon using time-dependent media. 
Instead of transmitting the wave through a spatial boundary we transmit it through an artificial temporal boundary, created by 
switching both parameters from constant to
dispersive with frequency.
We show that the resulting dynamics is sharply different from the spatial case, featuring both reflection and refraction in positive and negative regimes simultaneously.
We demonstrate our results analytically and numerically using electromagnetic medium.
In addition, we show that by a targeted dispersion tuning the temporal boundary can be made nonreflecting, while preserving both positive and negative refraction.

\end{abstract}

}

\maketitle

\IEEEdisplaynontitleabstractindextext

%
\IEEEpeerreviewmaketitle

\section{Introduction}

\IEEEPARstart{E}{xploring} wave propagation properties in time-varying media has attracted interest for several decades \cite{morgenthaler1958velocity,felsen1970wave,fante1971transmission,ruiz1978characteristics,aberg1995propagation,mendoncca2002time,xiao2014reflection,chen2019discrete}, and still remains in the scientific spotlight \cite{bruno2020negative,pacheco2020antireflection,victor2021brewster,castaldi2021exploiting,apffel2022experimental,galiffi2022photonics,yin2022floquet,pacheco2022time,li2022nonreciprocal}. The research spans continuous time dependence of the medium and of the excitation signal, abrupt time dependence also known as temporal boundary, temporal analogies of wave phenomena in propagation between spatially different media, and more.

One of the most basic analogies is the interaction of waves with boundaries.
When a wave of a certain frequency and momentum hits a spatial boundary, which physically separates two media with distinct properties, one transmitted and one reflected wave are created. Due to conservation of energy, the incident, transmitted and reflected waves have the same frequency but different momenta. 

The situation is entirely different for the case of a temporal boundary.
Such boundary is artificial, representing a sudden change in the constitutive parameters of a given medium, e.g. permittivity $\varepsilon$ and permeability $\mu$ in electromagnetics. A change that switches from constant $\varepsilon_1,\mu_1$ to other constant $\varepsilon_2,\mu_2$, splits a wave into two new waves, one propagating forward and the other backward, representing temporal refraction and reflection. Since in this case the energy is not conserved, all the waves keep the same momentum but have different frequencies  \cite{fante1971transmission,mendoncca2002time,xiao2014reflection,chen2019discrete}. 

Switching one of the parameters to dispersive in frequency was also studied \cite{fante1971transmission,solis2021time}, revealing splitting into more waves than two, depending on the dispersion order.
Methods to prevent the temporal reflection were suggested, including temporal analogy of impedance matching \cite{xiao2014reflection} and of impedance transformers \cite{pacheco2020antireflection,castaldi2021exploiting}.
 
A notable spatial wave propagation phenomenon is negative refraction of the so-called left-handed type \cite{ziolkowski2001wave,pendry2004negative,engheta2005positive}.
This phenomenon occurs when a wave is transmitted from a medium with constant and positive $\varepsilon,\mu$ into a medium with negative $\varepsilon,\mu$,
obtained when both parameters are frequency-dispersive and have an overlapping negative range. 
The phase and group velocities of the wave then become anti-parallel, which for one-dimensional propagation means negative phase velocity and positive group velocity. In higher spatial dimensions the phenomenon is also manifested by a negative angle of refraction.
Demonstration of negative refraction attracted an immense attention, and was realized in electromagnetics, acoustics and elasticity using metamaterials (architectured structures) \cite{smith2004metamaterials,lai2004composite,aydin2005observation,seo2012acoustic,liu2011elastic,sirota2019tunable,sirota2020active}, which created an effective medium with the required dispersion via modulation in space. 

In this work we are interested in obtaining the temporal analogue of negative refraction. In particular, we study the underlying conditions for this phenomenon to occur at a temporal boundary, and the fundamental differences from the spatial boundary scenario.

\section{Time-varying double-positive and double-negative index regions}

\begin{figure}
\begin{center}
    \includegraphics[width=7.9cm]{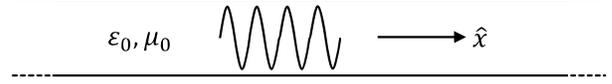} 
\end{center}
\caption{Schematic of the electromagnetic waveguide in medium I.}
\label{fig:Schematic_Diagram}
\end{figure}

The system we consider is an infinite electromagnetic waveguide characterized by the vacuum permittivity and permeability, $\varepsilon_0$ and $\mu_0$, as schematically illustrated in Fig. \ref{fig:Schematic_Diagram}. 
A signal of frequency $\omega_0$ propagating in the $\hat{x}$ direction governs the waveguide during the interval $t < t_s$, which we denote by medium I.
At $t=t_s$ we introduce a temporal boundary by changing the medium parameters in a step-like fashion, so that the permittivity and the permeability are modulated by
\begin{subequations} 
    \begin{align}
    \tilde{\varepsilon}(\omega)&=\frac{\omega^2-\omega_{\beta}^2}{\omega^2},
    \label{eq:Constitutive_Parameters_E} \\
    \tilde{\mu}(\omega)&=\frac{\omega^2-\omega_{m}^2}{\omega^2},
    \label{eq:Constitutive_Parameters_M}
    \end{align}
\end{subequations}
where $\omega_{\beta}=\beta\omega_{0}$ and $\omega_{m}=m\omega_{0}$ for constant and positive $m$ and $\beta$.
The relations in \eqref{eq:Constitutive_Parameters_E}-\eqref{eq:Constitutive_Parameters_M} are usually referred to as Drude model dispersion \cite{engheta2005positive}. In the general case, when $m\neq \beta$, this model is dubbed unmatched, whereas for $m=\beta$ the matched model is obtained.
We denote the interval $t\geq t_{s}$ by medium II, which is characterized by the dynamic permittivity and permeability $\varepsilon(\omega)=\varepsilon_0\tilde{\varepsilon}(\omega)$ and $\mu(\omega)=\mu_0\tilde{\mu}(\omega)$.
To obtain the field dynamics in medium II, we consider
Maxwell's equations in the time domain for general polarization fields, which read
\begin{subequations} 
    \begin{align}
    \nabla \times \textbf{E} &=-\mu_0\frac{\partial \textbf{H}}{\partial t}-\mu_0\frac{\partial \textbf{M}}{\partial t}, \label{eq:Constitutive_Equations_E} \\
    \nabla \times \textbf{H} &=\varepsilon_0\frac{\partial \textbf{E}}{\partial t}+\frac{\partial \textbf{P}}{\partial t}. 
    \label{eq:Constitutive_Equations_H}
    \end{align}
\end{subequations}
Here, $\textbf{E}=\hat{z}E(x,t)$ is a $\hat{z}$-directed electric field, $\textbf{H}=\hat{y}H(x,t)$ is a $\hat{y}$-directed magnetic field, and $\textbf{P}$ and $\textbf{M}$ are the polarization and magnetization fields, given by
\begin{subequations} 
    \begin{align}
    \textbf{P} &=\varepsilon_0(\tilde{\varepsilon}(\omega)-1)\textbf{E}\, \label{eq:P_field}, \\
    \textbf{M} &=(\tilde{\mu}(\omega)-1)\textbf{H}, \label{eq:M_field}
    \end{align}
\end{subequations}
where $\tilde{\varepsilon}(\omega)$ and $\tilde{\mu}(\omega)$ are defined in \eqref{eq:Constitutive_Parameters_E}-\eqref{eq:Constitutive_Parameters_M}. $\textbf{P}$ and $\textbf{M}$ exist only in medium II, since in medium I we have $\tilde{\varepsilon}=\tilde{\mu}=1$.
Combining \eqref{eq:Constitutive_Equations_E}-\eqref{eq:Constitutive_Equations_H} into a wave equation, transforming it to the complex domain $s$ using Laplace transform and employing \eqref{eq:P_field}-\eqref{eq:M_field}, we obtain
\begin{equation}
    \tilde{E}(x;s)=s\frac{(s^2+\omega_{m}^2)E(x,0^{-})+s\frac{\partial E}{\partial t}(x,0^{-})}{s^4+(\omega_{0}^2+\omega_{m}^2+\omega_{\beta}^2)s^2+\omega_{m}^2\omega_{\beta}^2}.
    \label{eq:Simp_Laplace_E}
\end{equation}
Setting the electric field in medium I to be $E(x,t<0)=\cos(\omega_{0}t-kx)$, where $k$ is the wavenumber,
the initial conditions for the electric field in medium II at $t=0^{-}$ become
\begin{subequations} \label{eq:IC_Gen}
    \begin{align}
        E(x,0^{-})&=\cos(kx),
        \label{eq:IC_E} \\
        \frac{\partial E}{\partial t}(x,0^{-})&=\omega_{0}\sin(kx).
        \label{eq:IC_E_DOT}
    \end{align}
\end{subequations}
When a wave is incident on a spatial boundary, the wavenumber $k$ is not preserved, while the frequency $\omega$ is, due to energy conservation. In contrast, when a wave is incident on a temporal boundary, the wavenumber $k$ is preserved, while the frequency $\omega$ is not, due to momentum conservation \cite{fante1971transmission,mendoncca2002time,xiao2014reflection,chen2019discrete}.
Therefore, we expect the waves generated in medium II by the temporal interface to have new frequencies, which are different from $\omega_0$ and satisfy the momentum conservation, i.e., $k_{II}=k_{I}=k$.
Solving the problem defined by \eqref{eq:Simp_Laplace_E}-\eqref{eq:IC_Gen} with the same $k$ in medium I and II, we obtain
\begin{align}
\begin{split}
    E(x,t)=&\sum_{i=1}^{4}a_{i}\cos(\omega_{i}t-kx)
            \\&=\frac{1}{2}\frac{\omega_1^2+\omega_0\omega_1-\omega_m^2}{\omega_1^2-\omega_3^2}\cos(\omega_1t-kx)
            \\&+\frac{1}{2}\frac{\omega_2^2+\omega_0\omega_2-\omega_m^2}{\omega_1^2-\omega_3^2}\cos(\omega_2t-kx)
            \\&+\frac{1}{2}\frac{\omega_m^2-\omega_0\omega_3-\omega_3^2}{\omega_1^2-\omega_3^2}\cos(\omega_3t-kx)
            \\&+\frac{1}{2}\frac{\omega_m^2-\omega_0\omega_4-\omega_4^2}{\omega_1^2-\omega_3^2}\cos(\omega_4t-kx).
    \label{eq:Ez_UDM_solution}
\end{split}
\end{align}
Here,
\begin{equation}
    \omega_{1-4}=\pm \frac{\omega_0}{\sqrt{2}}\sqrt{1+m^2+\beta^2 \pm \sqrt{(1+m^2+\beta^2)^2-4m^2\beta^2}}
    \label{eq:UDM_Freq_after_switch}
\end{equation}
are the frequencies generated by the temporal boundary, obtained from the solution of the characteristic equation $s^4+(\omega_{0}^2+\omega_{m}^2+\omega_{\beta}^2)s^2+\omega_{m}^2\omega_{\beta}^2=0$ in \eqref{eq:Simp_Laplace_E}. These frequencies satisfy $\omega_{2}=-\omega_{1},\; \omega_{4}=-\omega_{3}$, and the amplitudes $a_{i}$ are the reflection and transmission coefficients.
The solution in \eqref{eq:UDM_Freq_after_switch} includes four different frequencies. These frequencies are also the particular solutions of the general dispersion relation
\begin{equation}
    k=\frac{\omega}{c_0}\sqrt{\tilde{\mu}(\omega)\tilde{\varepsilon}(\omega)},
    \label{eq:Wavenumber}
\end{equation}
where $c_0=1/\sqrt{\mu_0\varepsilon_0}$ is the wave speed in medium I.
As part of this derivation, the matched Drude model case $m=\beta\equiv\alpha$ is of a particular interest. By substituting $\omega_m=\omega_{\beta}\equiv\omega_{\alpha}$, \eqref{eq:UDM_Freq_after_switch} simplifies to
\begin{equation}
    \omega_{1-4}=\pm \frac{\omega_0}{\sqrt{2}}\sqrt{1+2\alpha^2 \pm \sqrt{1+4\alpha^2}}.
    \label{eq:MDM_Freq_after_switch}
\end{equation}
On the other hand, substituting \eqref{eq:Constitutive_Parameters_E}-\eqref{eq:Constitutive_Parameters_M} in \eqref{eq:Wavenumber} and equating to $k_{I}=\omega_0/c_0$, results in
\begin{equation} 
    \omega\left(|\omega^2-\omega_{\alpha}^2|-\omega_0\omega\right)=0,
    \label{eq:MDM_Omega_eqn}
\end{equation}
which is satisfied by the frequencies in \eqref{eq:MDM_Freq_after_switch}. Since  \eqref{eq:MDM_Freq_after_switch} also implies that $\omega_{1,2}^2>\omega_{\alpha}^2$ and $\omega_{3,4}^2<\omega_{\alpha}^2$, the explicit form of \eqref{eq:MDM_Omega_eqn} reads
\begin{subequations} 
    \begin{align}
        \omega_{1}\left(\omega_{1}^2-\omega_0\omega_{1}-\omega_{\alpha}^2\right)=0,
        \label{eq:Dispersion_Result_Omega_1} \\
        \omega_{3}\left(\omega_{\alpha}^2-\omega_0\omega_{3}-\omega_{3}^2\right)=0.
        \label{eq:Dispersion_Result_Omega_3}
    \end{align}
\end{subequations}
It can be observed that \eqref{eq:Dispersion_Result_Omega_1} and \eqref{eq:Dispersion_Result_Omega_3} are exactly the numerators of the amplitudes $a_2$ and $a_3$ in \eqref{eq:Ez_UDM_solution}, meaning that both vanish in the matched dispersion case. The time domain solution \eqref{eq:Ez_UDM_solution} then takes the form
\begin{align}
\begin{split}
    E(x,t)=&+\frac{1}{2}\frac{\omega_1^2+\omega_0\omega_1-\omega_{\alpha}^2}{\omega_1^2-\omega_3^2}\cos(\omega_1t-kx)
            \\&+\frac{1}{2}\frac{\omega_{\alpha}^2-\omega_0\omega_4-\omega_4^2}{\omega_1^2-\omega_3^2}\cos(\omega_4t-kx).
    \label{eq:Ez_MDM_solution}
\end{split}
\end{align}
In the next section we analyze the kinematics and dynamics of the fields in \eqref{eq:Ez_UDM_solution} and \eqref{eq:Ez_MDM_solution}.

\section{Simultaneous negative and positive reflection and refraction}

\begin{figure}
\begin{center}
\begin{tabular}{c c}
    \textbf{(a)} \\ \includegraphics[height=5.8cm]{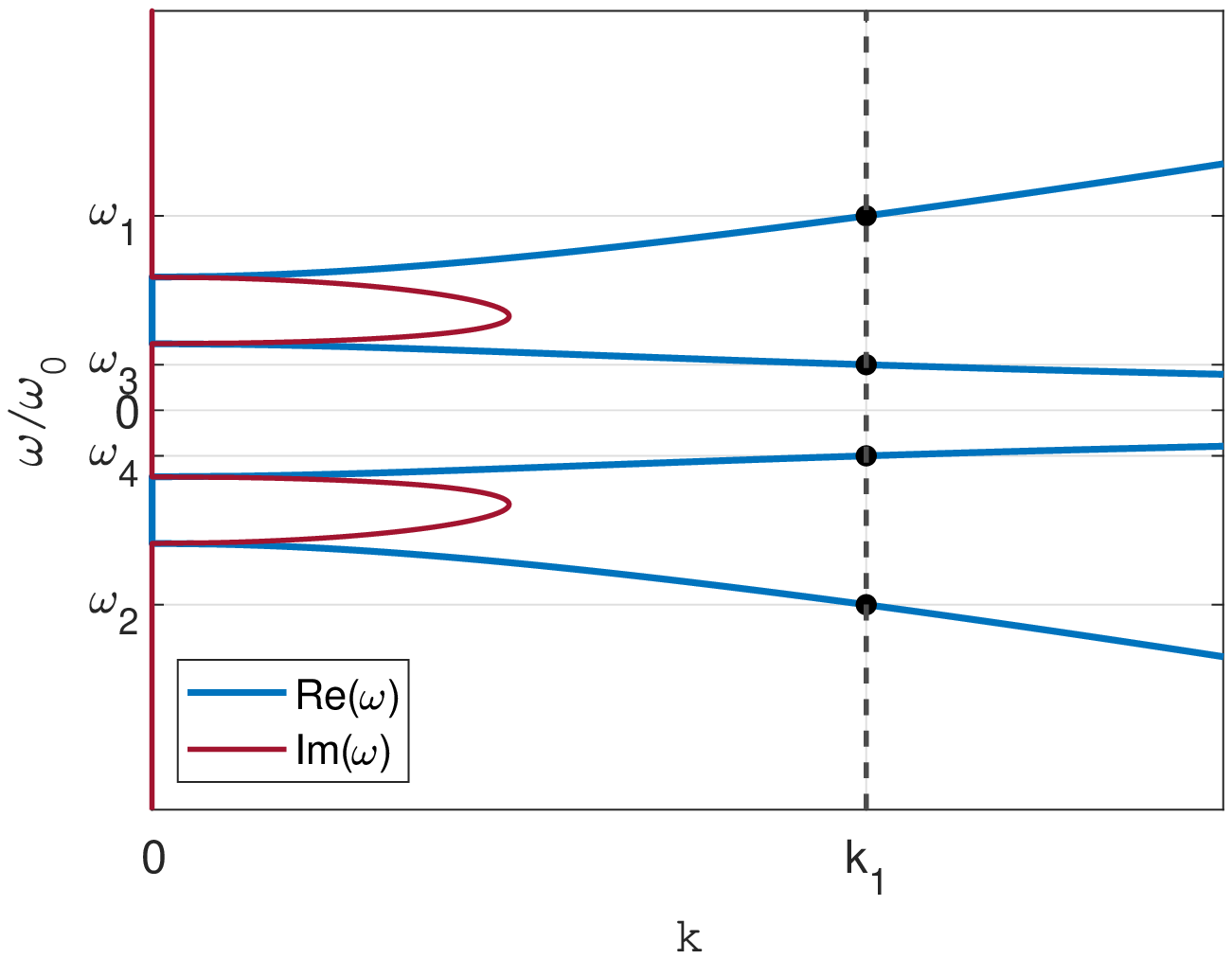} \\
    \textbf{(b)} \\ \includegraphics[height=5.8cm]{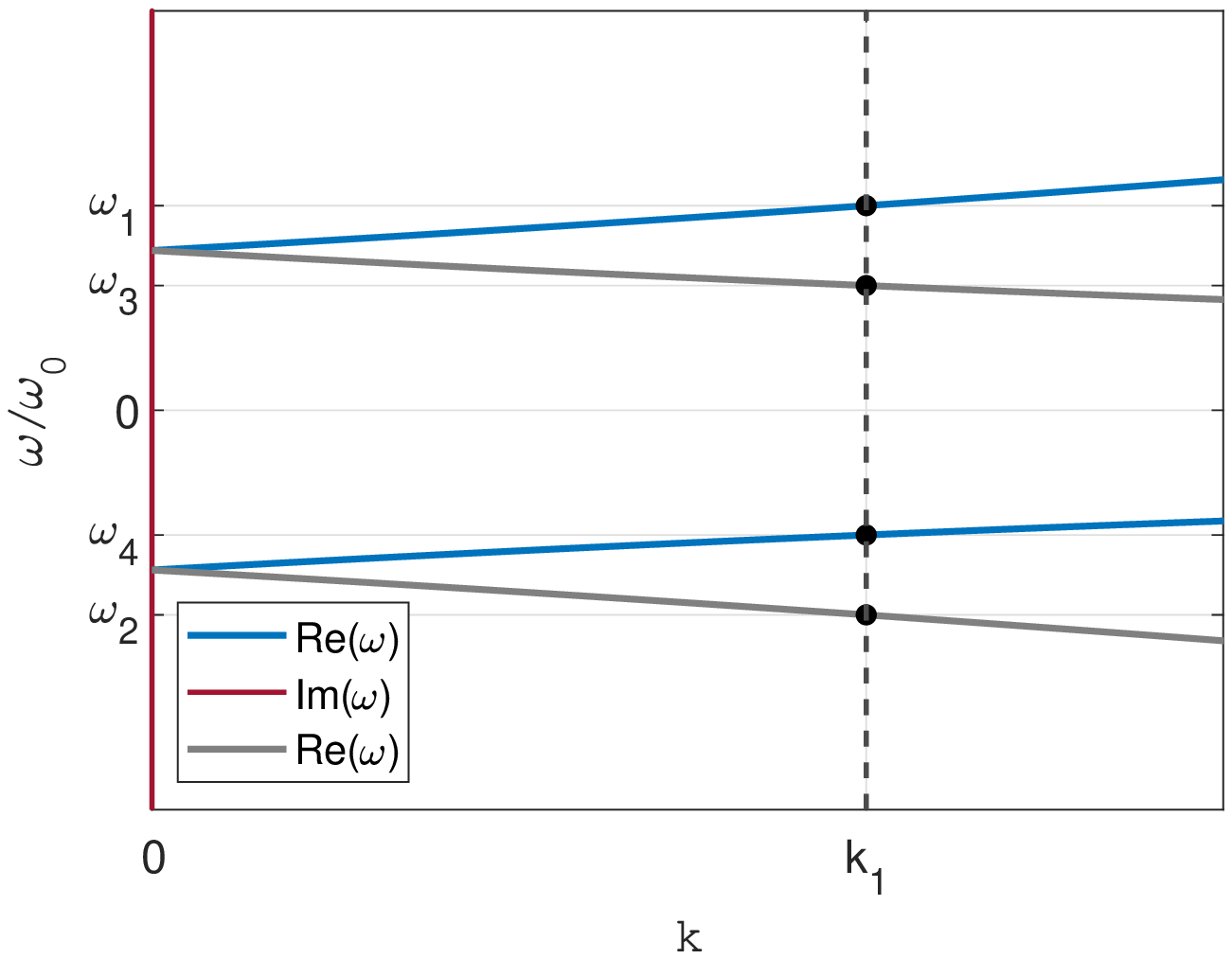}
    \end{tabular}
\end{center}
\caption{Dispersion diagram of medium II. Black dots indicate the four frequencies $\omega_{1-4}$ created by the temporal boundary. (a) The unmatched model. Both real (blue) and imaginary (red) spectrum exist, but $\omega_{1-4}$ are in the real region. (b) The matched model. Only real spectrum exists. Two of the four branches (gray) are cancelled.}
\label{fig:Dispersion_Diagram}
\end{figure}

Figure \ref{fig:Dispersion_Diagram} depicts the dispersion relation given by \eqref{eq:Wavenumber}.
In Fig. \ref{fig:Dispersion_Diagram}(a), which represents the unmatched dispersion case, the frequency has both real (blue) and imaginary (red) values, indicating waves propagating with constant and non-constant amplitudes, respectively. 
The black dots indicate the frequencies generated by the temporal boundary, $\omega_{1-4}$, given in \eqref{eq:UDM_Freq_after_switch}, which implies that the dynamics in medium II is confined to the propagation-only region.

In contrast, in the matched case, which is shown in Fig. \ref{fig:Dispersion_Diagram}(b), the frequencies can only have real values, which is manifested by the disappearance of the red curves.
Moreover, here, only two frequencies out of the four given in \eqref{eq:MDM_Freq_after_switch} are allowed in medium II due to the fact that the reflection coefficients of two waves out of the four have a zero amplitude. Therefore, only $\omega_{1}$ and $\omega_{4}$ are relevant.

\begin{figure}
\begin{center}
\begin{tabular}{c c}
    \textbf{(a)} \\ \includegraphics[height=2.3cm]{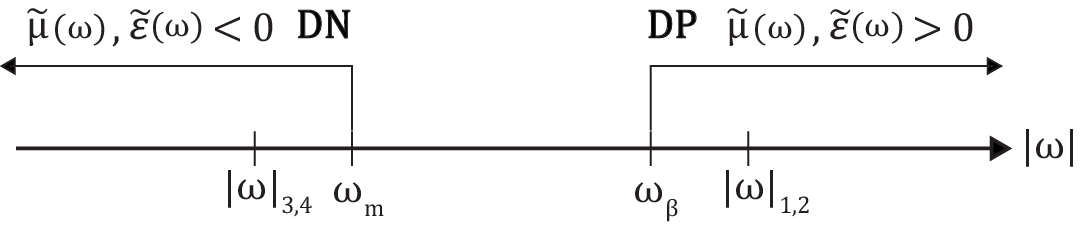} \\
    \textbf{(b)} \\ \includegraphics[height=2.3cm]{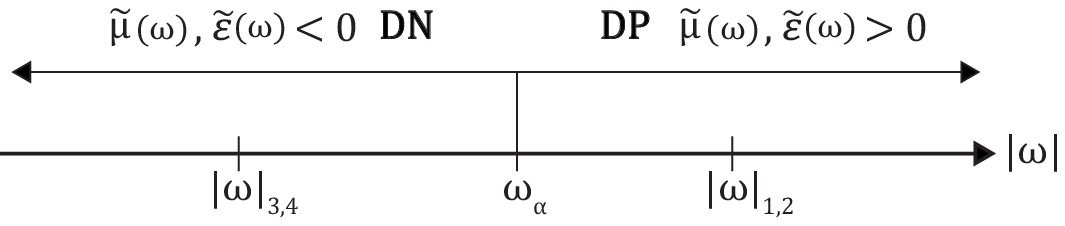}
    \end{tabular}
\end{center}
\caption{Schematic representation of the dynamical regimes with respect to frequency. Double-positive (DP) holds when both $\tilde{\varepsilon}(\omega)$ and $\tilde{\mu}(\omega)$ are positive, and double-negative (DN) when both are negative. (a) The matched dispersion model. (b) The unmatched dispersion model.}
\label{fig:Omega_Axis}
\end{figure}

We now discuss the resulting dynamical regimes.
Figure \ref{fig:Omega_Axis} depicts a schematic frequency axis, sectioned into regions that indicate the sign of the constitutive parameters in \eqref{eq:Constitutive_Parameters_E}-\eqref{eq:Constitutive_Parameters_M}, as well as the corresponding locations of the absolute values of the frequencies $\omega_{1-4}$.
For both unmatched and matched dispersion models, Fig. \ref{fig:Omega_Axis}(a) and \ref{fig:Omega_Axis}(b), $\omega_{1,2}$ belong to the region where $\tilde{\mu}(\omega)>0$ and $\tilde{\varepsilon}(\omega)>0$, indicating the double-positive (DP) regime, whereas $\omega_{3,4}$ belong to region where $\tilde{\mu}(\omega)<0$ and $\tilde{\varepsilon}(\omega)<0$, indicating the double-negative (DN) regime. 

The dynamical characteristics of these regimes can be directly interpreted from the dispersion diagrams in Fig. \ref{fig:Dispersion_Diagram}. First we consider the unmatched case. The waves corresponding to $\omega_{1}$ and $\omega_{2}$ propagate with the phase velocity $v_{ph_{1,2}}=\pm \omega_{1}/k$. 
Their group velocity, $v_{g}=\partial\omega/\partial k$, is of the same sign as the phase velocity, as evident from the positive slope of the dispersion curves at the intersection points with the momentum $k_1$.
The DP regime deduced for these waves from the frequency axis in Fig. \ref{fig:Omega_Axis}(a) is thus manifested by the parallel phase and group velocities.

The waves corresponding to $\omega_{3}$ and $\omega_{4}$ have a phase velocity of $v_{ph_{3,4}}=\pm \omega_{3}/k$. Now, however, the slope of the corresponding dispersion curves is negative, indicating group velocities that are anti-parallel to the phase velocities, thus manifesting the underlying DN regime. 
Since the waves with positive (negative) group velocities indicate the refracted (reflected) waves, the $\omega_1$ wave undergoes a positive refraction, i.e. propagating forward in terms of energy, and appearing as forward in terms of phase. Accordingly, the $\omega_4$ wave undergoes a negative refraction, i.e. propagating forward and appearing as backward. Remarkably, both positive and negative refraction appear simultaneously, unlike for interaction with a spatial boundary.
The $\omega_2$ and $\omega_3$ waves are both reflected, propagating backward, and appearing respectively as backward and forward, indicating positive and negative reflection.

In the matched case, the two existing waves of $\omega_{1}$ and $\omega_{4}$, have a positive group velocity, with a positive (negative) phase velocity for $\omega_{1}$ ($\omega_{4}$), corresponding to the DP (DN) regime deduced from Fig. \ref{fig:Omega_Axis}(b).
The response thus consists of refractive waves only, where one ($\omega_1$) is in a positive refraction, i.e. propagating forward and appearing forward, and the other ($\omega_4$) in a negative refraction, i.e. propagating forward but appearing backward. Therefore, in this case no actual reflection takes place.

It is interesting to note that the solution in which more than two waves arise from the temporal switch can be received also by changing only one constitutive parameter to dispersive \cite{solis2021time}. In that case, all waves are in a DP regime, with two waves positively refracting (with $v_g>0$ and $v_{ph}>0$) and the other two reflecting (with $v_g<0$ and $v_{ph}<0$). However, only when both parameters are changed, two different dynamical regimes, the DP and the DN, can be created, and temporal negative refraction can be supported.

\begin{figure*}[htpb]
\begin{center}
\setlength{\tabcolsep}{-2pt}
\begin{tabular}{c c c}
    \textbf{(a)} & \textbf{(c)} & \textbf{(e)} \\ \includegraphics[width=\x cm, valign=c]{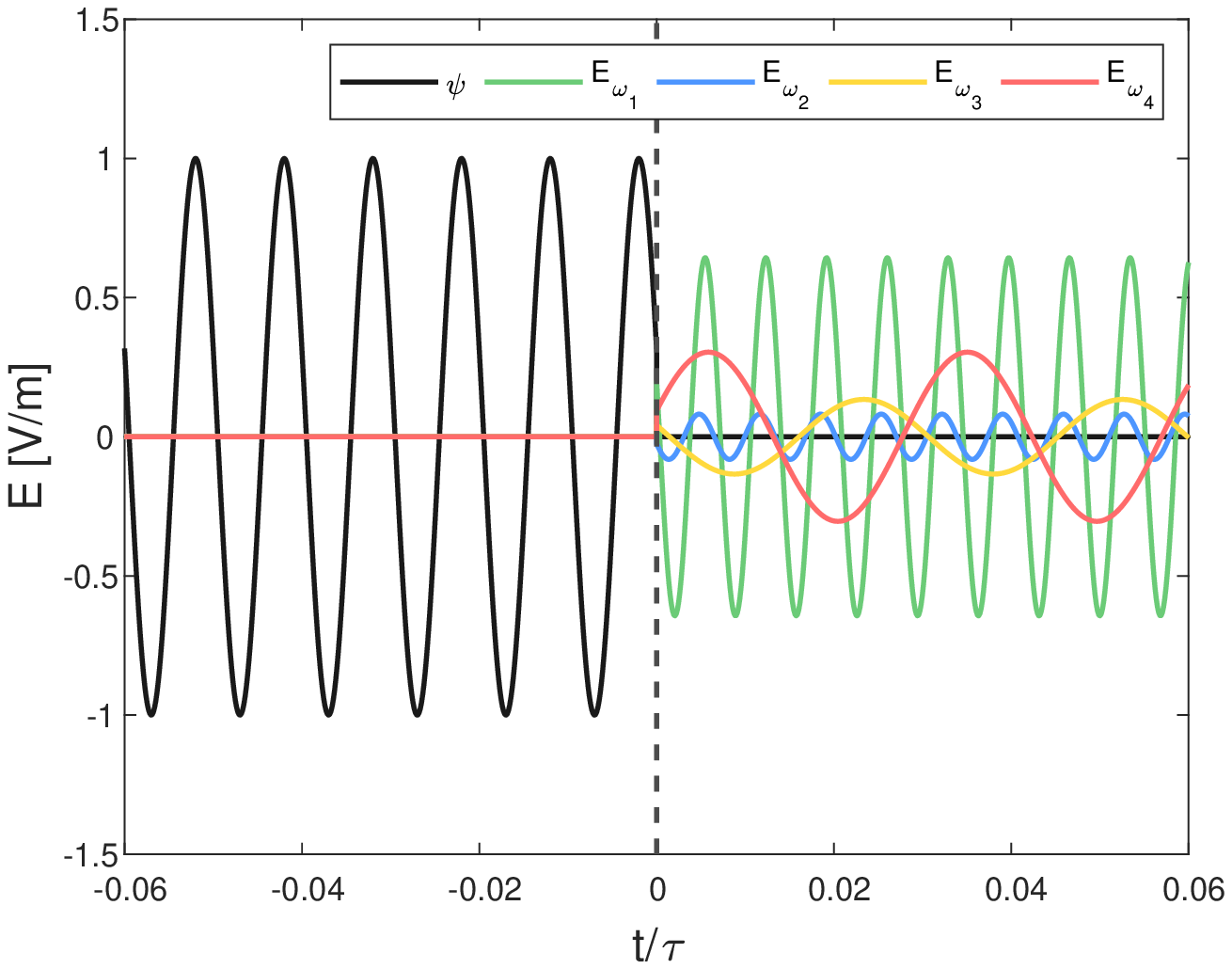} & \includegraphics[width=\x cm, valign=c]{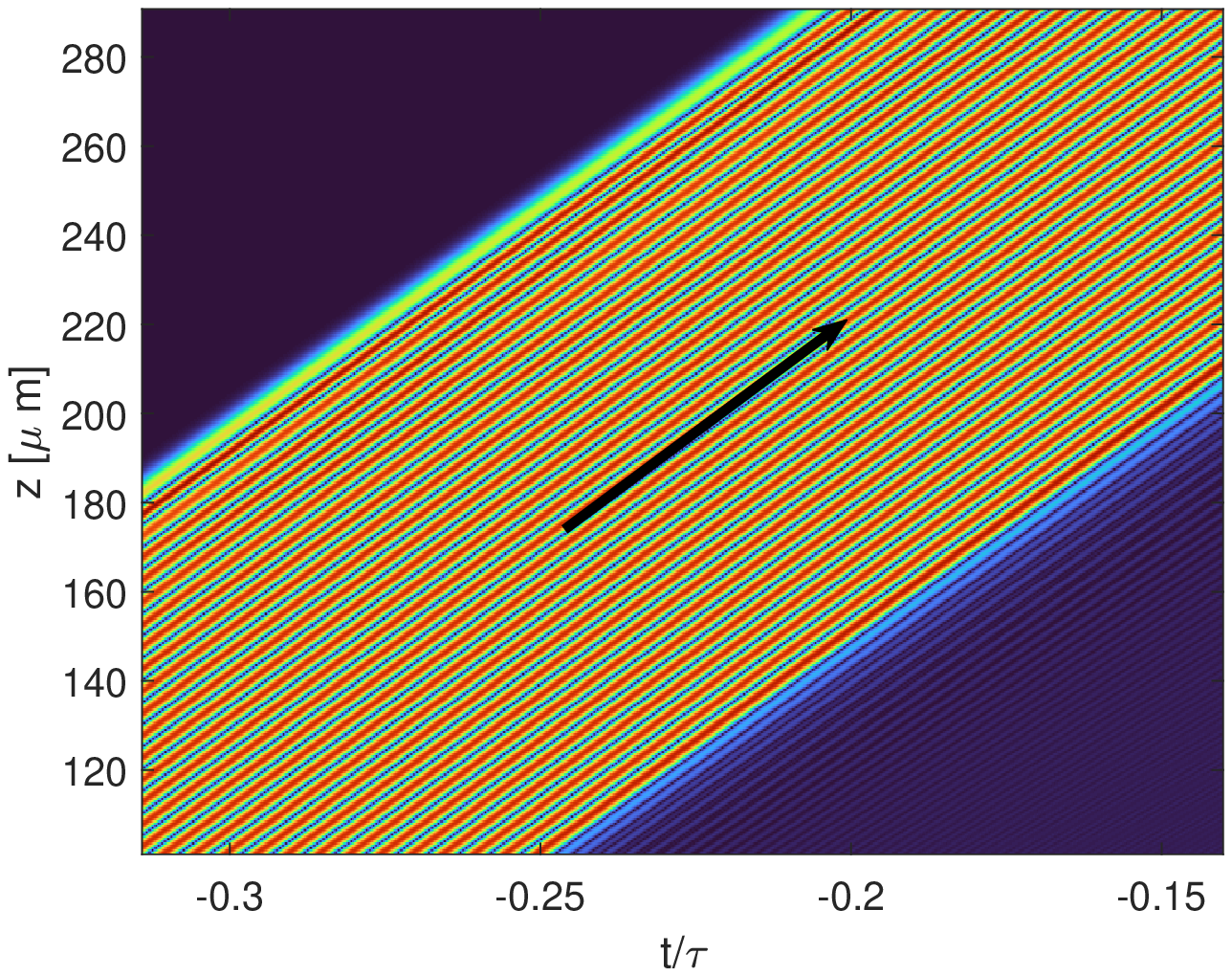} & \includegraphics[width=\x cm, valign=c]{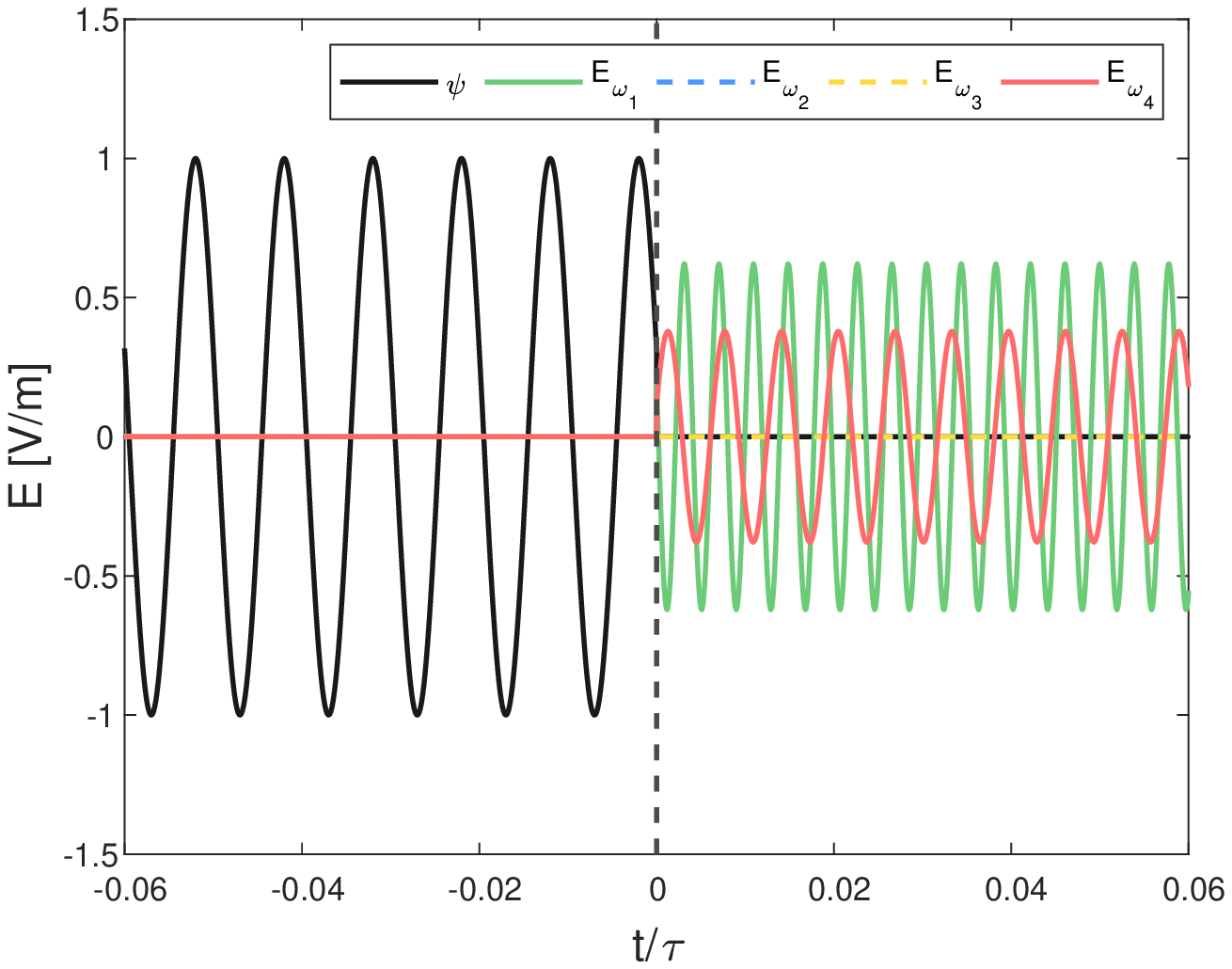} \\
    \textbf{(b)} & \textbf{(d)} & \textbf{(f)} \\ \includegraphics[width=\x cm, valign=c]{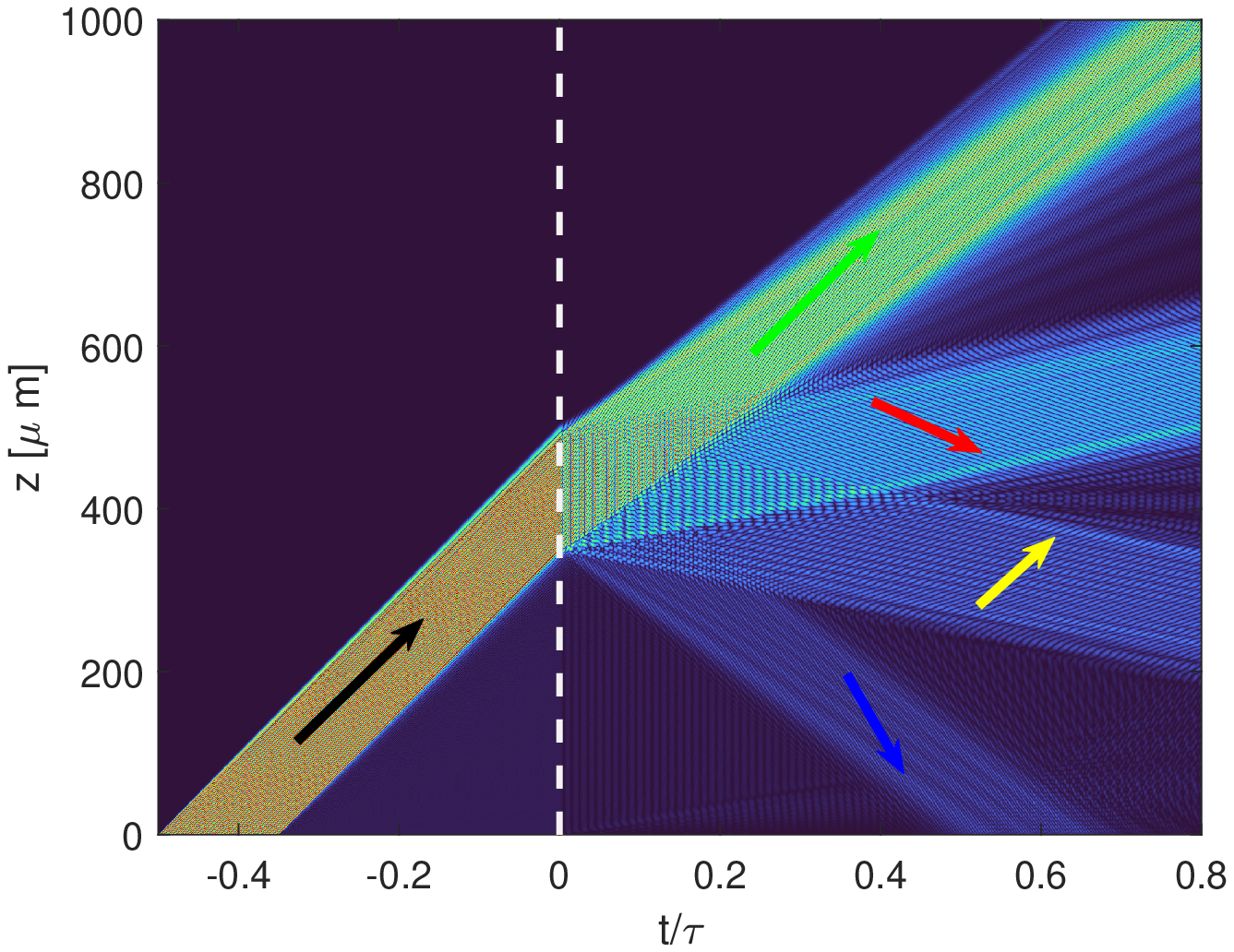} & \includegraphics[width=\x cm, valign=c]{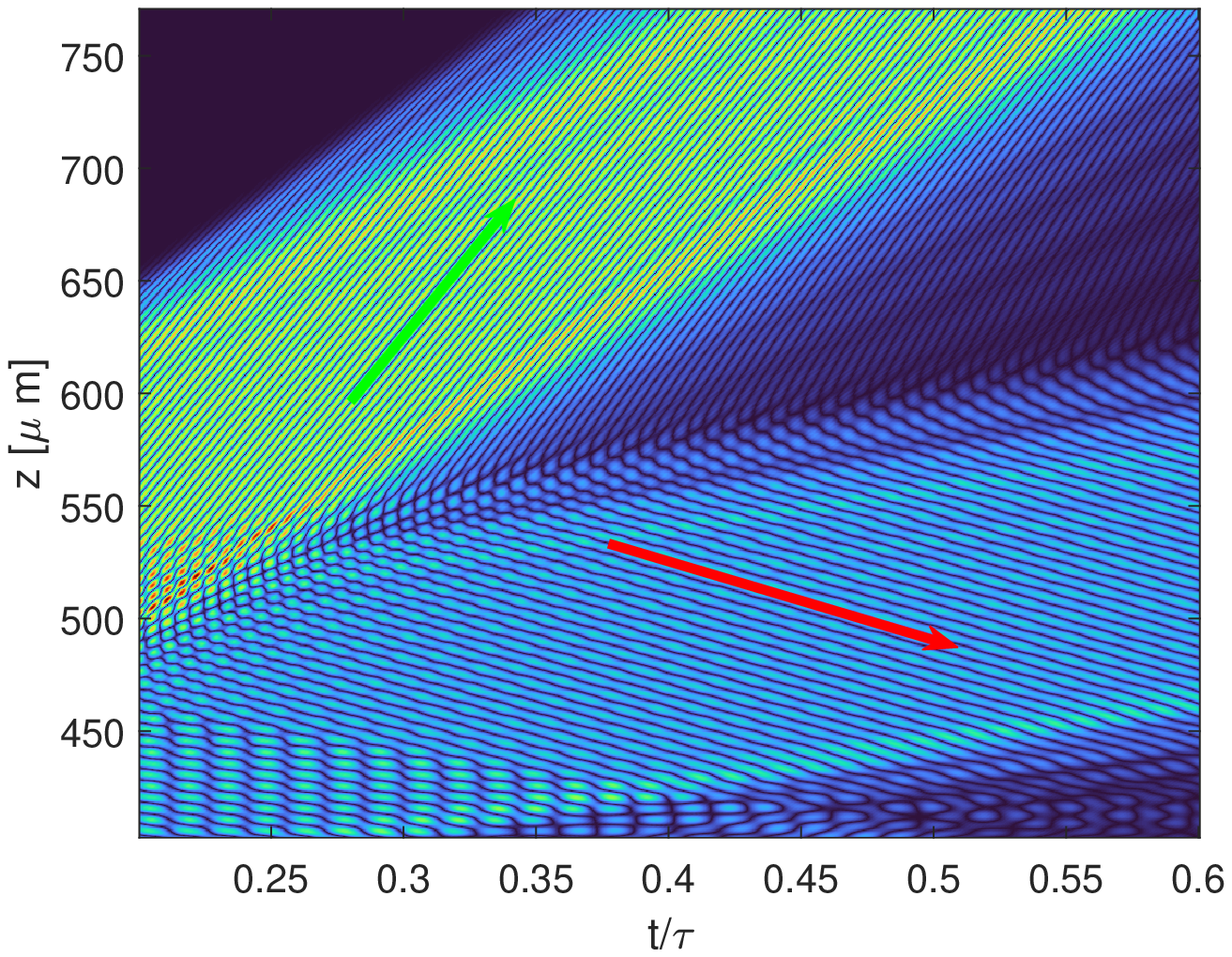} &
    \includegraphics[width=\x cm, valign=c]{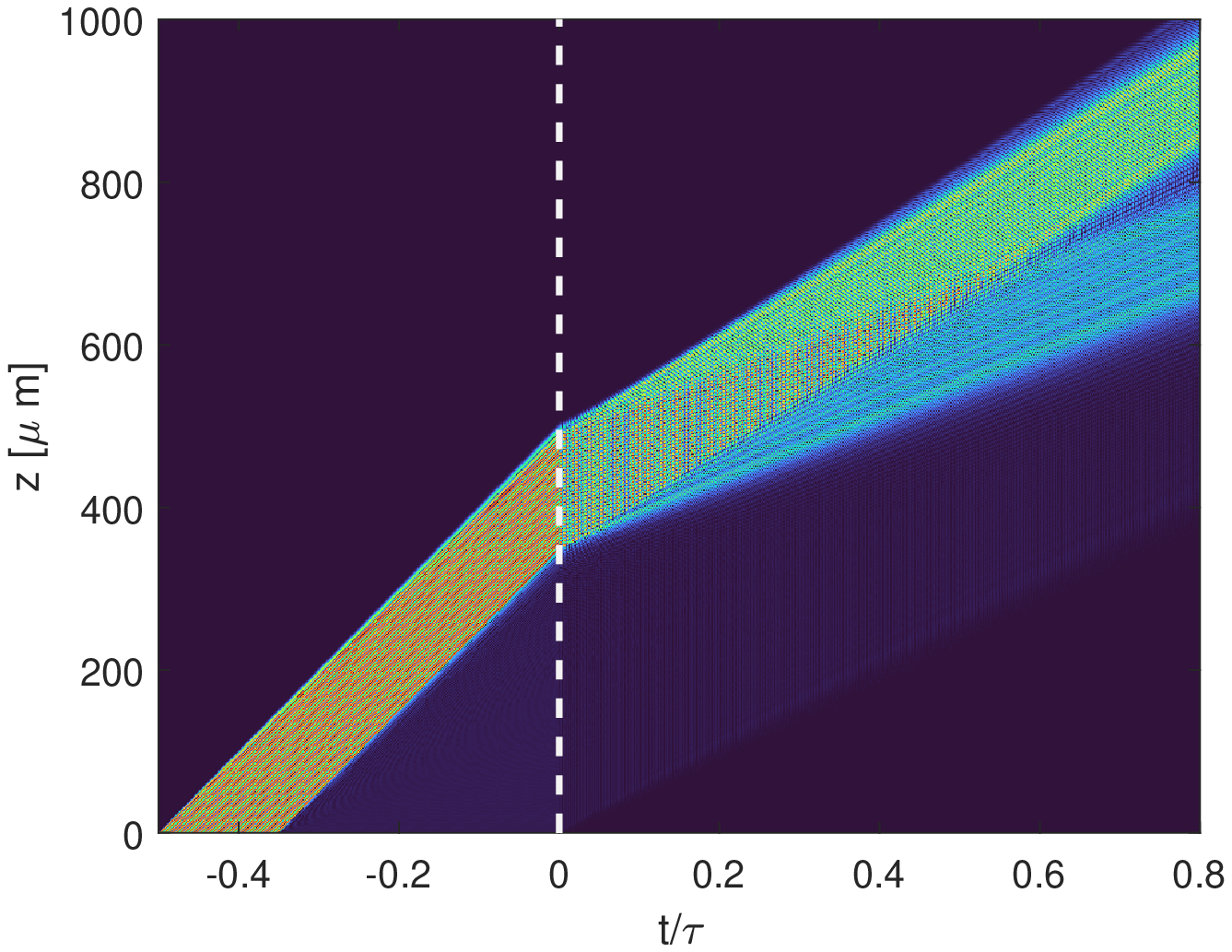}
    \end{tabular}
\end{center}
\caption{Simulations of the wave interaction with a temporal boundary between uniform and dispersive medium according to \eqref{eq:Constitutive_Parameters_E}-\eqref{eq:Constitutive_Parameters_M}. (a)-(d) The unmatched dispersion model. (a) Analytical solution, depicting a harmonic plane wave propagation of frequency $\omega_0=628$ kHz before hitting the boundary (black), and the four waves of frequencies $\omega_1=1.4604\omega_{0}$ (green), $\omega_2=-1.4604\omega_{0}$ (blue), $\omega_3=0.3424\omega_{0}$ (yellow) and $\omega_4=-0.3424\omega_{0}$ (red), after hitting the boundary, corresponding to \eqref{eq:Ez_UDM_solution}. (c) Numerical FDTD solution, depicting a finite-width burst modulated by $\omega_0$ and forming a beam-like propagation in space-time. After hitting the boundary the beam splits into four beams. The beams slopes represent the group velocities. The arrows slopes represent the phase velocities, whereas their colors correspond to the analytical waves in (a). (c),(d) Close-up on the phase fronts in the incident wave and the $\omega_1,\omega_4$ response waves. (e)-(f) The matched dispersion model. (e) Analytical solution, depicting the same incidence wave as in (a) before hitting the boundary (black), and the two waves of frequencies $\omega_1=2.561\omega_{0}$ (green), $\omega_4=-2.561\omega_{0}$ (red), after hitting the boundary, corresponding to \eqref{eq:Ez_MDM_solution}. (f) The corresponding numerical FDTD solution.}
\label{fig:Results}
\end{figure*}

We now demonstrate the actual time evolution of the electric field before and after its interaction with the temporal boundary, as given in Fig. \ref{fig:Results}.
Figures \ref{fig:Results}(a)-(d) address the unmatched case. Figure \ref{fig:Results}(a) presents the analytical solution in \eqref{eq:Ez_UDM_solution}. The incident signal, depicted in black, is a harmonic signal of frequency $\omega_{0}=628$ kHz, assumed to be propagating in vacuum. The resulting signals, shown separately, were obtained for $\beta=0.5$ and $m=1$ in \eqref{eq:UDM_Freq_after_switch}, leading to the frequencies $\omega_{1}=1.4604\omega_{0}$, $\omega_{2}=-1.4604\omega_{0}$, $\omega_{3}=0.3424\omega_{0}$ and $\omega_{4}=-0.3424\omega_{0}$, which are depicted in green, blue, yellow and red, respectively.

Figure \ref{fig:Results}(b) presents the numerical simulation of equations \eqref{eq:Constitutive_Equations_E}-\eqref{eq:Constitutive_Equations_H} and \eqref{eq:P_field}-\eqref{eq:M_field} using Finite Difference Time Domain (FDTD).
The underlying constitutive parameters \eqref{eq:Constitutive_Parameters_E}-\eqref{eq:Constitutive_Parameters_M} were implemented via higher order partial differential field equations, discretized in space and in time. We used a computational domain of length $L=1000 \mu m$, and the input frequency $\omega_0$ as in the analytical simulation.
Medium I is defined by the interval $-0.5\tau \leq t < 0$, where $\tau=L/c_{0}$ is the system time constant, and medium II by the interval $0 \leq t \leq 0.8\tau$.

The response is presented in a 2D space-time plot. The incident signal, which is a harmonicaly-modulated burst, thus forms a beam-like propagation.
When this burst hits the temporal boundary, it splits into four separate beams, starting at the dashed line.
The slopes of the beams correspond to the signs of the group velocities.
The slopes of the arrows depicted on top of the beams, which are aligned with the phasefronts, correspond to the phase velocities; positive for the green and yellow waves and negative for the red and blue waves.   
This is emphasized in the close-ups given in Fig. \ref{fig:Results}(c) for the incident wave, and in Fig. \ref{fig:Results}(d) for the refracted waves, which are red and green. 

Figures \ref{fig:Results}(e)-(f) address the matched case. The analytical solution \eqref{eq:Ez_MDM_solution}, calculated for $\alpha=2$ in \eqref{eq:MDM_Freq_after_switch} with $\omega_{1}=2.5616\omega_{0}$ and $\omega_{4}=-2.5616\omega_{0}$, is depicted in Fig. \ref{fig:Results}(e). It features the same source wave as in the unmatched case in black, and the refracted-only waves, green for the positive refraction (DP regime) wave of $\omega_{1}$, and red for the negative refraction (DN regime) wave of $\omega_{4}$.
Figure \ref{fig:Results}(f) depicts the corresponding numerical solution, featuring only two beams splitting out in medium II, as expected.

\section{Conclusion}

We demonstrated the phenomenon of negative refraction in a time-modulated medium, as a temporal analogue to the celebrated spatial effect. 
We applied a step-like time modulation to the permittivity and the permeability of an electromagnetic medium, converting from uniform values into frequency-dispersive values that follow the Drude model. 
As a result of the interaction with this temporal boundary, the source wave split into four waves, each having a different frequency. 
Two of the new frequency magnitudes fall within the range where both permittivity and permeability are positive, whereas the other two where both are negative, thus forming DP and DN index regimes, respectively. 

In the DP regime the waves have parallel group and phase velocities, either both positive or both negative, respectively indicating positive refraction and reflection.
In the DN regime phase and group velocities are anti-parallel for both waves. A positive (negative) group and a negative (positive) phase velocity indicate negative refraction (reflection).
The two reflected waves can be completely eliminated by using the matched Drude model.
Remarkably, at the temporal boundary both negative and positive refraction occur simultaneously, in a sharp contrast to a spatial boundary, at which refraction can be either positive or negative.


%


\bibliographystyle{IEEEtran}
\bibliography{references_TV}

\end{document}